\documentstyle[twocolumn,aps,floats,babel,spanish,prl,latexsym,amsmath,amssymb,eqsecnum]{revtex}

\begin{document}

\draft \tolerance = 1000

\setcounter{topnumber}{1}
\renewcommand{\topfraction}{1}
\renewcommand{\textfraction}{0}
\renewcommand{\floatpagefraction}{1}
\newcommand{\br}{{\bf r}}

\twocolumn[\hsize\textwidth\columnwidth\hsize\csname@twocolumnfalse\endcsname

\title{{\rule{17.5cm}{.15mm}{\bfseries{\\\vspace{3mm}Cosmological Models with ``Some" Variable Constants\\\rule{17.5cm}{.15mm}}}}}
\author{Jos\'e Antonio Belinch\'on}
\address{Grupo Inter-Universitario de An\'alisis Dimensional\\
Dept. F\'isica ETS Arquitectura  UPM\\
Av. Juan de Herrera 4 Madrid 28040 España\\
E-mail: jabelinchon@usa.net}
\maketitle

\begin{abstract}
{\em Various models are under consideration with metric type flat FRW i.e.
with k=0 whose energy-momentum tensor is described by a perfect fluid whose
generic equation state is $p=\omega \rho $ and taking into account the
conservation principle $div(T_{ij})=0$, but considering some of the
``constants'' as variable. A set of solutions through dimensional analysis
is trivially found. The numeric calculations carried out show that the
results obtained are not discordant with those presently observed for
cosmological parameters. However, the model seems irreconcilable with
electromagnetic and quantum quantities. This makes us think that we are
working with faulty hypothesis from the start}.\medskip

\end{abstract}
\vspace{.4cm}

]
\section{\bf Introduction.}

In a recent paper (see \cite{TONY1}) a study was carried out of the
behaviour of $G$ and $\Lambda $ ``constants'' in the framework described by
a cosmological model with flat FRW symmetries and whose moment-energy tensor
was described by a perfect fluid whose state equation is : $(p=\omega \rho $
/ $\omega =const.)$ taking into account the conservation principle. In light
of coincidences encountered with O'Hanlon and Tam's work (see \cite{OHA}) it
was decided to demonstrate that we could also reach the same results as
those of the authors in the case of a model describing a Universe with
predominance of matter. With the hypothesis carried out in previous work
(see \cite{TONY1}) i.e. the only real constants taken into consideration are
the speed of light $c$ and the Boltzmann constant $k_B.$ In this work the
``constants'' $G,\hbar ,a,e,m_i$ and $\Lambda $ are considered as scalar
functions dependent on time showing that one of our models reaches the same
results of those of O'Hanlon and Tam. It should also be pointed out how the
use of Dimensional Analysis (DA) enables us to find in a trivial way, a set
of solutions to this type of models with $k=0$ y taking into account the
conservation principle, as it will be seen that the different equations
posed are not trivially integrable, but ``without going overboard'' as the
method has certain limitations. The numerical calculations carried out show
that the results obtained are not discordant with those presently observed
for cosmological parameters, however, the model seems irreconcilable with
electromagnetic and quantum quantities. The idea of being able to unite in
one field both gravitation and the rest of the forces has been, since the
beginning of the century, a very active field of work. The inconsistencies
observed in our model makes us think ``{\em momentaneously}'' that the
approach to the problem is inadequate, that is to say, that we are working
with faulty hypothesis from the start.\medskip\ 

The paper is organized as follows: In the second section the equations of
the model are presented and some small considerations on the dimensional
method followed are made. In the third section, use is made of dimensional
analysis to obtain a solution to the main quantities appearing in the model.
In the fourth section two specific models are studied - one with radiation
predominance and the other with matter predominance and finally in the fifth
section a brief summary and concise conclusions are made.

\section{\bf The model.}

The modified field equations are as follows:
\begin{equation}
\label{e0}R_{ij}-\frac 12g_{ij}R-\Lambda (t)g_{ij}=\frac{8\pi G(t)}{c^4}%
T_{ij}
\end{equation}
and it is imposed that: 
$$
div(T_{ij})=0 
$$
where $\Lambda (t)$ represent the cosmological ``constant''. The basic
ingredient of the model are:

\begin{enumerate}
\item  The line element defined by: 
$$
ds^2=-c^2dt^2  
+f^2(t)\left[ \frac{dr^2}{1-kr^2}+r^2\left( d\theta ^2+\sin {}^2\theta d\phi
^2\right) \right] 
$$
we only consider here the case $k=0.$

\item  The energy-momentum tensor defined by: 
$$
T_{ij}=(\rho +p)u_iu_j-pg_{ij}\qquad \qquad p=\omega \rho  
$$
where $\omega $ is a numerical constant such that $\omega \in \left[
0,1\right] $
\end{enumerate}

Whit this supposition the equation that govern the model are as follows: 
\begin{equation}
\label{e1}2\frac{f\,^{\prime \prime }}{f\,}+\frac{(f\,^{\prime })^2}{f\,^2}%
=- \frac{8\pi G(t)}{c^2}p+c^2\Lambda (t)\ \ 
\end{equation}
\begin{equation}
\label{e2}3\frac{(f\,^{\prime })^2}{f\,^2}=\frac{8\pi G(t)}{\,c^2}\rho
+c^2\Lambda (t)\qquad \quad \ 
\end{equation}
\begin{equation}
\label{e3}div(T_{ij})=0\text{ }\Leftrightarrow \rho ^{\prime }+3(\omega
+1)\rho \frac{f^{\prime }}f=0 
\end{equation}
integrating equation (\ref{e3}) it is obtained the well-known relationship. 
\begin{equation}
\label{e4}\rho =A_\omega f^{-3(\omega +1)} 
\end{equation}
where $f$ represent the scale factor that appear in the metric and $A_\omega 
$ is the constant of integration that has different dimensions and physical
meaning depending on the state equation imposed i.e. depends on $\omega $.

Following the Kalligas et al`s work (see \cite{K}), if we derive equation (%
\ref{e2}) and it is simplified with (\ref{e1}) it is obtained the
relationship: 
\begin{equation}
\label{f1}G\rho ^{\prime }+3(1+\omega )\rho G\frac{f^{\prime }}f+\rho
G^{\prime }+\frac{\Lambda ^{\prime }c^4}{8\pi }=0
\end{equation}

From equations (\ref{f1}) and (\ref{e3}) we obtain the next equation that
relate $G$ with $\Lambda $%
\begin{equation}
\label{f2}G^{\prime }=-\frac{\Lambda ^{\prime }c^4}{8\pi \rho }
\end{equation}
from all these relationship it is obtained the following differential
equation that it is not immediately integrated (see \cite{K}): 
\begin{equation}
\label{f4}\frac{\rho ^{\prime }\rho ^{\prime \prime }}{\rho ^2}-\left( \frac{%
\rho ^{\prime }}\rho \right) ^3=12\pi (\omega +1)^2\frac{G\rho ^{\prime }}{%
c^2}
\end{equation}
for this reason we utilize the dimensional method. This equation also we can
integrate through similarity and dimensional method following a well-
established way to integrate pde and odes. This last option is studied in
other paper (see \cite{T0}). In this case we work a naive Dimensional
Analysis.\medskip\ 

The followed dimensional method needs to make these distinctions. It is
necessary to know beforehand the set of fundamental quantities together with
one of the unavoidable constant (in the nomenclature of Barenblatt
designated as governing parameters). In this case the only fundamental
quantity is the cosmic time $t$ as can be easily deduced from the
homogeneity and isotropy supposed for the model. The set of unavoidable
constant are in this case the speed of light $c,$ the integration constant $%
A_\omega $ (obtained from equation (\ref{e4}) that depending on the state
equation will have different dimensions and physical meaning) and the
Boltzman constant $k_B$ that will be taking into account to relate
thermodynamics quantities

In a previous paper (see \cite{T}) the dimensional base was calculated for this
type of models, being this $B=\left\{ L,M,T,\theta \right\} $ where $\theta $
stands for dimensions of temperature. The dimensional equations of each of
the governing parameters is: 
$$
\left[ t\right] =T\quad \left[ c\right] =LT^{-1}\quad \left[ A_\omega
\right] =L^{2+3\omega }MT^{-2}
$$
$$ 
\left[ k_B\right] =L^2MT^{-2}\theta ^{-1} 
$$

All the derived quantities will be calculated in function of these governing
parameters, that is say, in function of the cosmic time $t$ and the set of
unavoidable constants $c,$ $k_B$ and $A_\omega $ with respect to the
dimensional base $B=\left\{ L,M,T,\theta \right\} .$

\section{{\bf {Solutions through D.A.}}}

We are going to calculate through dimensional analysis D.A. i.e. applying
the Pi Theorem, the variation of $G(t)$ in function on $t$ and temperature $%
\theta ,$ $G(\theta )$ (see \cite{Z}), the Planck's constant $\hbar (t),$the
radiation constant $a(t),$ the charge of the electron $e(t),$ the mass of an
elementary particle $m_i (t)$\text{, }the variation of the cosmological
``constant'' $\Lambda (t),$ the energy density $\rho (t),$ the matter
density $\rho _m(t),$ the radius of the universe $f(t),$ the temperature $%
\theta (t)$, the entropy $S(t)$ and finally the entropy density $s(t)$

\subsection{{\bf Calculation of }${\bf G(t)}${\bf .}}

As we have indicated above, we are going to accomplish the calculation of
the variation of $G$ applying the Pi theorem. The quantities that we
consider are: $G=G(t,c,A_\omega ).$ with respect to the dimensional base $%
B=\left\{ L,M,T,\theta \right\} .$ We know that $\left[ G\right]
=L^3M^{-1}T^{-2}$%

Through a direct aplication of Pi Theorem we obtain a single monomial that leads to the following expression for $G$%
\begin{equation}
\label{r1}G(t)\propto \frac{t^{1+3\omega }c^{5+3\omega }}{A_\omega } 
\end{equation}

If we want to relate $G$ with $\theta $ (see \cite{Z}) the solution that DA
give us is: $G=G(t,c,A_\omega ,a,\theta ).$ We need to introduce a new
dimensional ``constant'' $a$, in this case thermodynamics, to relate the
temperature whit the rest of quantities. The same result is obtained if
we consider $k_B.$%
\begin{equation}
\label{r2}G(\theta )\propto A_\omega ^{\frac 1{3(\omega +1)}}c^4\left(
a\theta ^4\right) ^{\frac{\omega -1}{3(\omega +1)}} 
\end{equation}

\subsection{\bf Calculation of the Planck's constant $\hbar (t):$}

$\hbar =\hbar $ $(t,c,A_\omega )$ where its dimensional equation is $\left[
\hbar \right] =L^2MT^{-1}$%
\begin{equation}
\label{r3}\hbar (t)\propto A_\omega c^{-3\omega }t^{1-3\omega } 
\end{equation}

\subsection{{\bf Calculation of the radiation ``constant'' }${\bf a(t)}${\bf :}}

$a=a(t,c,A_\omega ,k_B)$ where its dimensional equation is $\left[ a\right]
=L^{-1}MT^{-2}\theta ^{-4}$ 
\begin{equation}
\label{r4}k_B^{-4}a(t)\propto A_\omega ^{-3}c^{9\omega -3}t^{9\omega -3} 
\end{equation}

\subsection{{\bf Calculation of the electron charge }${\bf e(t)}:$}

$e=e(t,c,A_\omega ,\epsilon _0)$ where its dimensional equation is $\left[
e^2\epsilon _0^{-1}\right] =L^3MT^{-2}$ 
\begin{equation}
\label{r5}e^2(t)\epsilon _0^{-1}\propto A_\omega c^{1-3\omega }t^{1-3\omega
} 
\end{equation}

\subsection{\bf Calculation of the mass of an elementary particle $m_i (t):$}

$m_i=m_i(t,c,A_\omega )$ where its dimensional equation is $\left[
m_i\right] =M$ 
\begin{equation}
\label{r6}m_i(t)\propto A_\omega c^{-2-3\omega }t^{-3\omega } 
\end{equation}

\subsection{\bf Calculation of the cosmological ``constant'' $\Lambda (t).$}

$\Lambda =\Lambda (t,c,A_\omega )$ where $\left[ \Lambda \right] =L^{-2}$%
\begin{equation}
\label{r7}\Lambda (t)\propto \frac 1{c^2t^2} 
\end{equation}
it is observed that not depends on $A_\omega $ i.e. it is not depends on
state equation. This solution will be valid for both models.

\subsection{\bf Calculation of the energy density $\rho (t)$}

$\rho =\rho (t,c,A_\omega )$ with respect to the base $B$ its dimensional
equation is: $\left[ \rho \right] =L^{-1}MT^{-2}$%
\begin{equation}
\label{r8}\rho (t)\propto A_\omega \left( ct\right) ^{-3(\omega +1)} 
\end{equation}

\subsection{\bf Calculation of the radius of the Universe $f(t).$}

$f=f(t,c,A_\omega )$ where its dimensional equation is $\left[ f\right]
=L\Longrightarrow $%
\begin{equation}
\label{r9}f(t)\propto ct 
\end{equation}
it is observed that no depends on $A_\omega $ i.e. is not depend on state
equation. This solution is valid for both models. Then: 
$$
q=-\frac{f^{\prime \prime }f}{\left( f^{\prime }\right) ^2}=0 
$$
$$
H=\frac{f^{\prime }}f=\frac 1t 
$$
$$
d_H=ct\lim _{t_0\rightarrow 0}\int_{t_0}^t\frac{dt^{\prime }}{f(t^{\prime })}%
=\infty 
$$
i.e. there is no horizon problem, since $d_H$ diverge when $t_0\rightarrow
0. $

\subsection{\bf Calculation of the temperature $\theta (t).$}

$\theta =\theta (t,c,A_{\omega ,}a)$ where $\left[ \theta \right] =\theta
\Longrightarrow $%
\begin{equation}
\label{r10}a^{\frac 14}\theta (t)\propto A_\omega ^{\frac 14}\left(
ct\right) ^{-\frac 34(1+\omega )}
\end{equation}
we can too calculate it in function of $k_B$ i.e. $\theta =\theta
(t,c,A_{\omega ,}k_B)$ 
\begin{equation}
\label{r11}k_B\theta (t)\propto A_\omega c^{-3\omega }t^{-3\omega }
\end{equation}
we may check that this relationship is verified:%
$$
\rho =a\theta ^4=A_\omega (ct)^{-3(\omega +1)}=A_\omega (f)^{-3(\omega +1)} 
$$

\subsection{{\bf Calculation of the entropy }${\bf S}(t).$}

$S=S(c,A_{\omega ,}a)$ where $\left[ S\right] =L^2MT^{-2}\theta ^{-1}$ 
\begin{equation}
\label{r12}S(t)\propto \left( A_\omega ^3a(ct)^{3(1-3\omega )}\right)
^{\frac 14} 
\end{equation}

\subsection{{\bf Entropy density }${\bf s}(t).$}

$s=s(t,c,A_{\omega ,}a)$ where $\left[ s\right] =L^{-1}MT^{-2}\theta ^{-1}$ 
\begin{equation}
\label{r13}s(t)\propto \left( A_\omega ^3a\right) ^{\frac 14}\left(
ct\right) ^{\frac 94(1+\omega )} 
\end{equation}

\section{\bf Different Cases.}

All the following cases can be calculated without difficulty. Two specific
models are studied: in first place a universe with radiation predominance
which corresponds to the imposition of $\omega =1/3$ in the state equation;
and in second place a model describing a universe with matter predominance
corresponding to the imposition of $\omega =0$ as state equation.

\subsection{\bf Model with radiation predominance}

In this case the behaviour of the ``constants'' obtained is the following: 
 
$$
G(t)\propto A_\omega ^{-1}t^2c^6 \hspace{1cm} G\propto t^2 
$$
$$
G\propto c^4\left( A_\omega a\right) ^{-\frac 12}\theta ^{-2}\hspace{1cm} 
G\propto \theta ^{-2} 
$$
$$
\hbar \propto A_\omega c^{-1}t^0 \hspace{1cm} \hbar \propto const. 
$$
$$
a\propto k_B^4A_\omega ^{-3}c^0t^0\qquad \hspace{1cm} a\propto const.
$$
$$
e^2\epsilon _0^{-1}\propto A_\omega c^0t^0 \hspace{1cm} e^2\epsilon
_0^{-1}\propto const. 
$$
$$
m_i\propto A_\omega c^{-3}t^{-1} \hspace{1cm} m_i\propto t^{-1} 
$$
$$
\Lambda \propto c^{-2}t^{-2} \hspace{1cm} \Lambda \propto t^{-2} 
$$

While the result obtained for the rest of quantities is: 
$$
f\propto ct \hspace{1cm} f\propto t 
$$
$$
\rho \propto A_\omega \left( ct\right) ^{-4} \hspace{1cm} \rho \propto t^{-4} 
$$
$$ 
\theta \propto k_B^{-1}A_\omega c^{-1}t^{-1}\hspace{1cm} \theta \propto t^{-1}
$$
$$
S\propto \left( A_\omega ^3a\right) ^{\frac 14}\hspace{1cm} S\propto const. 
$$
$$
s\propto \left( A_\omega ^3a\right) ^{\frac 14}\left( ct\right) ^{-3} \hspace{1cm} s\propto t^{-3}
$$

In the first place it should be pointed out that with regard to the values
obtained for $G,\Lambda ,f,\rho $ and $\theta $ the same were obtained as
those already found in literature (see \cite{K},\cite{A},\cite{BER} and \cite
{ABD}) demonstrating in this way that DA is a good tool for dealing with
these types of problems. In the same way the result obtained for $G(\theta
)\propto \theta ^{-2}$ coincides with that obtained by Zee (see \cite{Z}).
It is proven, amazingly, that the result obtained for the remainder of the
``constants'' is that these are constant in the model in spite of
considering them as variable, with the exception of the mass of an elemental
particle which varies as $m_i\propto t^{-1}$. Observe that with regards to
the ``indissoluble'' relationship $e^2\epsilon _0^{-1}\propto const.$ it can
be said that $e^2\propto \epsilon _0$ in such a way that the product $%
e^2\epsilon _0^{-1}\propto const.$ remains constant. If Moller and
Landau et al `s observations (see \cite{MOLL}) are taken into account, in
which the following relation $\epsilon _0\approx f(t)$ is shown , in our
case $\epsilon _0\approx f(t)\propto t$ we therefore find that $e^2\propto
\epsilon _0\propto t$ of the relation $c^2=(1/\epsilon _0\mu _0)$ we obtain $%
\epsilon _0\propto \mu _0^{-1}$. In the same way the following coincidences
can be observed: $\hbar \propto A_\omega c^{-1}$being and $a\propto \frac{%
k_B^4}{c^3h^3}$ if substituted the expression obtained though D.A. i.e. $%
a\propto k_B^4A_\omega ^{-3}$ can be recovered and results consistent. On
the other hand, from the relation $e^2\epsilon _0^{-1}\propto A_\omega $ and 
$\hbar \propto A_\omega c^{-1}$ it can be seen that $e^2\epsilon
_0^{-1}\propto \hbar c$ a relation known by all. All these results are
coherent with the behaviour of all energies, as $E=k_B\theta \propto t^{-1}$
, $E=mc^2\propto t^{-1}$if not this relation would be constant !`! $E=\hbar
\gamma \propto t^{-1}$ and the total Borh energy $E_{TB}=\frac{me^4}{%
\epsilon _0^2\hbar ^2}\propto t^{-1}.$\medskip \ 

It will now be checked if the results obtained are compatible with the
observational data available.\medskip \ 

From the equation (\ref{e4}) the value of the constant $A_\omega $ is
obtained (as in this model $\omega =1/3$ thus the denomination henceforth
will be as $A_1$). It is known that $\rho \approx 10^{-13.379}Jm^{-3}$ and $%
f\approx 10^{28}m$ with this data $A_1\approx 10^{100.5}m^3kgs^{-2}$ is
obtained. With this value of $A_1$ it is checked whether the value of $G$
predicted by our model is obtained. As $G(t)\propto A_\omega ^{-1}t^2c^6$
where $c\approx 10^{8.47}ms^{-1}$ and $t\approx 10^{20}s$%
$$
G(t)\propto A_\omega ^{-1}t^2c^6\approx 10^{-10.17}m^3kg^{-1}s^{-2} 
$$
i.e. our model is capable of recovering the value presently accepted of the
``{\em constant}'' $G$. If we proceed in the same way with the formula for $%
G(\theta )$ a value for $G$ about $G\approx
10^{-9.562}m^3kg^{-1}s^{-2}$ is obtained i.e. a little below that presently
observed. With regards to the cosmological ``constant'' it is observed that
: if $t\approx 10^{20}s$ and $c\approx 10^{8.4}ms^{-1}$ $\Rightarrow \Lambda
\approx 10^{-56}m^{-2}$ which corresponds to that presently accepted.\medskip%
\ 

If the value obtained for $A_1$ is taken into account it can be seen with
ease that the following value of the cosmic background radiation temperature
is obtained i.e $\theta \approx 10^{0.4361}K$ {\em if} the ``constant'' $a$
takes a value of $a\approx 10^{-15.1211}Jm^{-3}K^{-4}$ in the expression $%
a^{\frac 14}\theta (t)\propto A_\omega ^{\frac 14}\left( ct\right) ^{-1}$
i.e. we can also deduce through this result the value presently accepted of
the cosmic background radiation temperature also recovering the expression
for energy density $\rho =a\theta ^4$. Finally it should be pointed out that
our model is without the nominated problem of horizon although it is not yet
rid of the problem of entropy, also constant here.\medskip\ 

For the moment , we can see that the model works well ({\em fantastic}) but
we shall now see how it functions with respect to the electromagnetic and
quantum constants. With the values calculated previously we observe, much to
our disappointment, that we do not obtain (with the expressions indicated)
any of the values presently accepted for each of these ``constants''. For
example we see that $\hbar \propto A_1c^{-1}\approx 10^{91.5}Js^{-1\text{ }} 
$while for the radiation constant $a\propto k_B^4A_\omega ^{-3}\approx
10^{-300}$ $m^{-1}kgs^{-2}K^{-4}$ and $e^2\epsilon _0^{-1}\propto A_1\approx
10^{100.5}$ $m^3kgs^{-2}$ i.e. we are obtaining totally ``{\em preposterous
results}''. We can see that we are unable to reconcile our results with the
present values of the said ``constants''.\medskip\ 

Let us think now in a different way: from the relation $e^2\epsilon
_0^{-1}\propto A_1$ we obtain the value of the constant $A_1$which we shall
now call $A_1^{\prime }$ to avoid as far as possible confusion through
excessive notation. The value of this new constant is in the region of $%
A_1^{\prime }\approx 10^{-26}m^3kgs^{-2}$.so with this value of $A_1^{\prime
}$ we can recover the present values both of $\hbar $ and $a$ but none of
the cosmological parameters such as $G,\rho $ etc. how strange.\medskip\ 

We can see that even if by one path we can perfectly describe the
cosmological parameters $G,$ $f,$ $\rho ,$ $\theta $ and $\Lambda $ we
cannot recover the values of $\hbar ,$ $a,$ $e,$ $\epsilon _0$ etc. and vice
versa. This makes us think that this approach (I can now dare to qualify it
as simplistic as it is indeed a ``toy model'') is not correct and that
previous hypothesis should be taken into account or perhaps create an
adequate theoretical framework capable of describing both worlds ... ( I
think we have rediscovered America!).\medskip\ 

However, the approach displayed here is not totally wild for the following
reasons: {\em J.A. Wheeler} (see \cite{WHE}) stated that if the constants of
Physics must vary, these would do so in function of universal time. This
time is our universal time function which we can define in our model as we
have built it through a FRW metric type i.e. that our ST space-time can be
foliated in 3-spaces and these are different from one another in the value
of slice-labeling i.e. $trK$. The uranian mine in Gabon has given the
evidence necessary to be able to affirm that the masses and charges of
particles have changed with time, but which time? Proper or universal? One
can suddenly think that we are talking about proper time as changes of
masses and charges are proper ones! The collapse syndrome puts a limit in
the region of $10^{-37}/year$ on the variation of the charge of the
electron $e$ (with respect to proper time). A more thorough analysis of this
last section shows a series of difficulties ``the collapse syndrome''
hindered by Pauli's exclusion principle. The idea expressed by Wheeler is,
therefore, that the ``constants'' vary with respect to universal time and
not proper time. This argument sets universal time in a place of privilege
in the argument about the ``change'' in microphysics on cosmological scales.
On the other hand, as previously indicated, Moller and Landau et al
(see \cite{MOLL})) have shown a relation, now firmly established , by which
$\epsilon _0$ should vary according to the radius of the universe, $\epsilon _0\approx
f(t)$ .\medskip\ 

For all these reasons we have decided to carry out a similar study (i.e. it
seems our suppositions are not completely preposterous) however, the results
obtained surprise us.

\subsection{\bf Model with matter predominance.}

{\bf \ }In this case the behaviour of the ``constants'' obtained is the
following: 
$$
G(t)\propto A_\omega ^{-1}c^5t \hspace{1cm} G\propto t 
$$
$$
\hbar \propto A_\omega c^0t \hspace{1cm} \hbar \propto t 
$$
$$ 
a\propto k_B^4A_\omega ^{-3}c^{-3}t^{-3}\qquad \hspace{1cm} a\propto t^{-3} 
$$
$$ 
e^2\epsilon _0^{-1}\propto A_\omega ct \hspace{1cm} e^2\epsilon _0^{-1}\propto t
$$
$$
m_i\propto A_\omega c^{-2}t^0 \hspace{1cm} m_i\propto const. 
$$
$$
\Lambda \propto c^{-2}t^{-2} \hspace{1cm} \Lambda \propto t^{-2} 
$$

While the result obtained for the rest of quantities is: 
$$
f\propto ct \hspace{1cm} f\propto t 
$$
$$
\rho \propto A_\omega \left( ct\right) ^{-3} \hspace{1cm} \rho \propto t^{-3} 
$$
$$ 
\theta \propto k_B^{-1}A_\omega c^0t^0 \hspace{1cm} \theta \propto const. 
$$
$$ 
S\propto \left( A_\omega ^3a(ct)^3\right) ^{\frac 14} \hspace{1cm} S\propto t^3
$$
$$
s\propto \left( A_\omega ^3a\right) ^{\frac 14}\left( ct\right) ^{-9/4} \hspace{1cm} s\propto t^{-9/4} 
$$

In the same way as for the previous model it is seen that with respect to
quantities $G,$ $f,$ $\rho $ and $\Lambda $ the same results found in
literature (see \cite{K},\cite{BER} and \cite{ABD}) are obtained. With
regards to the rest of ``constants'' studied we see that in this case they
do vary.\medskip\ 

In particular regarding $G,$ $\hbar ,$ $e$ and $\rho $ the same results as
O'Hanlon et al (see \cite{OHA}) are obtained. These authors set out from the
Dirac model , its LNH, and by means of some pertinent modifications five
dimensionless numbers are obtained, in the same way as that of the Dirac
model, but this time reaching totally different results. With their five
dimensionless numbers together with the hypothesis that the mass of the
universe is constant (we do not need to make a similar hypothesis, the model
tells us $m_i\propto const.$) they are brought to the only way in which $G,$ 
$e^2$ and $\hbar $ vary. The $\rho $ average density of the universe mass
varies thus $\rho \propto t^{-3\text{ }}$ while $G\propto t$, $e^2\propto t$
and $\hbar \propto t$ while energy is conserved. This models responds, at
the same time, to the axioms of Milne's Kinematic Relativity (see \cite{M}).%
\medskip\ 

If $\rho $ is considered as mass density then $A_\omega $ (constant denoted
by $A_0$) $\left[ A_0\right] =M$ represents the universe mass, and the
expression $G(t)$ remains thus:%
$$
G\propto A_\omega c^3t\qquad G\propto t 
$$
verifying the Sciama formula $\rho Gt^2\approx 1$ (on inertia). Furthermore
if we take into account the numeric values of the constant and the quantity $%
t$ we obtain the present value of $G\approx 10^{-10.1757}m^3kg^{-1}s^{-2}$
i.e. $t\approx 10^{20}s,$ $c\approx 10^{8.5}ms^{-1}$ and $A_0\approx
10^{56}kg.$ This result was already obtained by Milne in 1935 (see \cite{M}%
). In the same way as in the previous case we are capable (with the value
obtained of $A_0)$ of recovering all the cosmological quantities but not
those corresponding to electromagnetism and the Planck ``constant''. And vice
versa. With regards to the cosmological ``constant'' in the same way as in
the previous case we see that if $t\approx 10^{20}s$ and $c\approx
10^{8.4}ms^{-1}$ $\Rightarrow \Lambda \approx 10^{-56}m^{-2}$ a value which
corresponds to that presently accepted.

\section{\bf Summary and conclusions.}

We have resolved through Dimensional Analysis DA a flat FRW model i.e. with $%
k=0$ whose energy-momentum tensor is described by a perfect fluid and taking
into account the conservation principle for said tensor i.e. $div(T_{ij})=0$
in which some constants are considered as variable i.e. as scalar functions
dependent on time. It has been proven that the dimensional technique used
resolves in a trivial manner the problem posed and that the results reached
correspond to those already existing in literature. New solutions have been
contributed as our model is more general as the variation of the
``constants'' $\hbar ,$ $e,$ $a,$ $\epsilon _0$ , $\mu _0$ and $m_i$ is
contemplated.\medskip

In the two models studied it is proven that the solutions obtained are
coherent with reference to cosmological parameters while for electromagnetic
and quantum quantities our model is not capable of adjusting itself to data
presently accepted for these. ``{\em At this time''} we believe that the
approach is erroneous, an unsettling question under revision.\medskip 

With regards to the model with matter predominance it is seen that it is
capable of theoretically justifying the O'Hanlon et al model, as we obtain
their same results without having to resort to any assumption or precise
numerological coincidence, even if it is based on the Dirac hypothesis.\

\end{document}